\documentclass[%
 aip,
 amsmath,amssymb,
 reprint,%
]{revtex4-1}

\usepackage{graphicx}
\usepackage{dcolumn}
\usepackage{bm}

\usepackage[utf8]{inputenc}
\usepackage[T1]{fontenc}
\usepackage{mathptmx}
\usepackage{etoolbox}

\makeatletter
\def\@email#1#2{%
 \endgroup
 \patchcmd{\titleblock@produce}
  {\frontmatter@RRAPformat}
  {\frontmatter@RRAPformat{\produce@RRAP{*#1\href{mailto:#2}{#2}}}\frontmatter@RRAPformat}
  {}{}
}%
\makeatother
\begin{document}

\preprint{AIP/123-QED}

\title{Universality classes of Anderson localization transitions in disordered three-dimensional non-Hermitian systems with exceptional points}
\author{C. Wang}
\affiliation{Center for Joint Quantum Studies and Department of Physics, School of Science, Tianjin University, Tianjin 300350, China}
\email{physcwang@tju.edu.cn}

\author{X. R. Wang}
\email{phxwan@ust.hk}
\affiliation{School of Science and Engineering, Chinese University of Hong Kong (Shenzhen), Shenzhen 518172, China}%

\date{\today}

\begin{abstract}
We conduct a numerical study of wave localization in disordered three-dimensional non-Hermitian systems featuring exceptional points. The energy spectrum of a disordered non-Hermitian Hamiltonian, exhibiting both parity-time and parity-particle-hole symmetries, forms a cross in the complex energy plane, with an exceptional point fixed at the origin. Near the exceptional point, the system experiences a disorder-driven quantum phase transition from extended to localized states, characterized as an Anderson localization transition in non-Hermitian systems. Notably, we identify a universal critical exponent that remains independent of the distribution of random variables. The model also supports Anderson localization transitions away from the exceptional points, albeit with different critical exponents. Furthermore, we investigate wave localization in a non-Hermitian system lacking parity-time symmetry, revealing distinct universality classes. By comparing the obtained critical exponents with those reported in the literature, we conclude that the presence of exceptional points introduces new universality classes that extend beyond the established 38-fold symmetry classification for non-Hermitian systems.
\end{abstract}

\maketitle

\section{Introduction}
\label{sec1}

Exceptional points (EPs) in non-Hermitian systems have garnered significant attention due to their unique spectral and dynamical properties. EPs are branch points of the complex eigenvalue spectrum at which two or more eigenvalues and their corresponding eigenvectors coalesce~\cite{mvBerry_book_2004}. Recent advancements in research have enhanced our understanding of the topology and robustness of EPs, revealing their potential to improve sensing capabilities~\cite{wj_Chen_nature_2017,hHodaei_nature_2017,hLoughlin_prl_2024,pDjorwe_prr_2024}, control wave propagation~\cite{maMiri_science_2019,hGangaraj_prl_2018,aYulaev_naturenano_2022}, and implement novel modes of energy transfer~\cite{hXu_nature_2016,wyang_prl_2025}. Experimental studies across various fields, including optics~\cite{sbLee_prl_2009,lFeng_op_2014,yhLai_nature_2019}, acoustics~\cite{czShi_nc_2016,vAchilleos_prb_2017,wwZhu_prl_2018,xWang_prl_2019}, and electronic systems~\cite{tStehmann_jpa_2004,zXiao_prl_2019} have successfully demonstrated the practical realization and manipulation of EPs. Simultaneously, theoretical investigations continue to explore the role of EPs in phase transitions~\cite{tSatoshi_pra_2016,cwang_prl_2018,hRyo_prr_2020,xDong_pra_2021,sAlexander_pra_2022,kDeng_prb_2023}, symmetry breaking~\cite{ykLu_sb_2018,mSakhdari_prl_2019,jhPark_nphys_2020,mIpsita_prl_2021,jhPark_nphys_2020}, and system stability~\cite{yZhiyenbayer_pra_2019,dPierre_prl_2021,fAlexander_prr_2025}.
Overall, these advancements underscore EPs as fundamental features with promising applications, further enriching our understanding of non-Hermitian physics.
\par

Among the various issues related to non-Hermitian systems, the disorder-induced localization problem is particularly significant, as disorder inevitably exists in all real materials~\cite{yHuang_prb_2020,cWang_prb_2020,kKohei_prl_2021,xlLuo_prl_2022,cWang_prb_2022,cWang_prb_2023,wWang_prl_2025,cWang_prl_2025,bLi_prl_2025,wJin_prl_2025}.
EPs lead to unique localization behaviors that contrast with those observed in traditional Hermitian models. Understanding the delocalization-localization transitions associated with EPs can reveal new phases and critical phenomena that are specific to non-Hermitian systems~\cite{cWang_prb_2022,cWang_prb_2023}. These insights have broad implications for a variety of practical applications, including optical systems~\cite{maMiri_science_2019,lFeng_op_2014}, quantum sensing~\cite{mZhang_prl_2019,jWiersig_nc_2020}, and topological materials~\cite{cDembowski_prl_2001,wHu_prb_2017,kKawabata_prl_2019,qSong_science_2021}.
\par

Most studies on localization properties of non-Hermitian systems featuring EPs focus on low dimensions, specifically one- or two-dimensional models~\cite{mAlvarez_prb_2018,cWang_prb_2022,cWang_prb_2023,smZhang_cpl_2024,dkHe_prb_2025}. While these low-dimensional systems have provided valuable insights into the fundamental mechanisms of non-Hermitian localization and the effects of EPs on spectral properties, it is essential to extend these investigations to higher-dimensional systems. Dimensionality plays a significant role in localization phenomena. Higher-dimensional systems can exhibit more complex behaviors, including distinct universality classes and critical exponents that may be absent or oversimplified in lower dimensions~\cite{pLee_rmp_1985,bKramer_rpp_1993,eFerdinand_rmp_2008}. Thus, exploring localization in higher-dimensional non-Hermitian systems around EPs is crucial for achieving a comprehensive understanding of their underlying physics and is also vital for potential applications in real-world, multi-dimensional platforms.
\par

In this study, we investigate the universality of Anderson localization transitions (ALTs) in a non-Hermitian model with EPs in three dimensions (3D). By introducing specific symmetries known as parity-time and parity-particle-hole symmetries~\cite{bCarl_prl_1998,cWang_prb_2023}, our non-Hermitian model is designed to exhibit EPs fixed at a particular location on the complex plane. As the degree of disorder increases, we observe that states in close proximity to the EP undergo ALTs. Through finite-size scaling analysis, we determine the critical exponent \( \nu=1.77 \) for these transitions. This finding demonstrates the concept of universality, indicating that the critical exponent remains invariant across different types of disorder. When moving away from the EP, the system's states can also experience delocalization-localization transitions; however, the universality of these transitions is altered. Furthermore, when the system's symmetries are modified, the EPs disappear, leading to different critical behaviors in the non-Hermitian system. Our results reveal that the presence of EPs significantly influences the critical characteristics of ALTs in non-Hermitian systems and predict the emergence of new universality classes.
\par

The paper is organized as follows. In Sec.~\ref{sec2}, we provide a brief introduction to the symmetry constraints necessary for realizing an EP at a fixed point in non-Hermitian systems. This section also includes the corresponding tight-binding model and a finite-size scaling analysis used to study the ALTs. The numerical results for identifying the universality classes of these systems are presented in Sec.~\ref{sec3}. Section~\ref{sec4} discusses the validity of conventional symmetry classification in the context of non-Hermitian systems with EPs. Finally, we conclude the paper in Sec.~\ref{sec5}.
\par

\section{Models and methods}
\label{sec2}

\subsection{Parity-time and parity-particle-hole symmetries}
\label{sec2_1}

Since ALTs occur in disordered systems, it is essential to incorporate randomness into our non-Hermitian model. The randomness may lead to variations in the positions of EPs across different random samples, making them challenging to locate precisely~\cite{cWang_prb_2023}. To address this issue, we introduce specific symmetries that constrain the system's energy spectrum, ensuring that the positions of the EPs remain invariant despite the presence of randomness. Below, we briefly outline the symmetry constraints applied in our model.
\par

One standard approach to generate an EP in a non-Hermitian system is to impose parity-time (PT) symmetry, which combines parity and time-reversal symmetries~\cite{bCarl_prl_1998}. Consider a single-particle Hamiltonian in the real space denoted as \( H=\sum_{\bm{ij}} c^\dagger_{\bm{i}} H_{\bm{ij}} c_{\bm{j}} \) with \( c^\dagger_{\bm{i}} \) and \( c_{\bm{i}} \) being the creation and annihilation operators on a lattice site \( \bm{i} \). Within the 38-fold symmetry classification for the non-Hermitian systems, time-reversal symmetry is defined as~\cite{kKohei_prx_2019} 
\begin{equation}
    \begin{gathered}
    U_{\mathcal{T}} H^\ast U^{-1}_{\mathcal{T}}=H
    \end{gathered}\label{eq_1_1}
\end{equation}
with \( U_{\mathcal{T}} \) being a unitary operator. If \( H \) is in the clean limit, it can be blocked-diagonalized as \( H=\sum_{\bm{k}} a^\dagger_{\bm{k}} h(\bm{k}) a_{\bm{k}} \) with \( h(\bm{k}) \) being the Bloch Hamiltonian. For Bloch Hamiltonians, time-reversal symmetry is defined as
\begin{equation}
    \begin{gathered}
    u_{\mathcal{T}} h^\ast(-\bm{k}) u^{-1}_{\mathcal{T}}=h(\bm{k}),
    \end{gathered}\label{eq_1_2}
\end{equation}
where \( u_{\mathcal{T}} \) is a unitary matrix. We also consider whether the Hamiltonian is invariant under parity inversion. Parity inversion symmetry is defined as~\citep{cWang_prb_2023}
\begin{equation}
    \begin{gathered}
    (U_{\mathcal{P}}\mathcal{P}) H (U_{\mathcal{P}}\mathcal{P})^{-1} =H
    \end{gathered}\label{eq_1_3}
\end{equation}
in the real space with the operator \( \mathcal{P} \) representing the spatial inversion that changes a lattice site \( \bm{i}=(i_x,i_y,i_z) \) to \( -\bm{i} \) and \( U_{\mathcal{P}} \) being a unitary operator. In the lattice-momentum space, parity inversion symmetry can be written as
\begin{equation}
    \begin{gathered}
    u_{\mathcal{P}} h(-\bm{k}) u^{-1}_{\mathcal{P}}=h(\bm{k}),
    \end{gathered}\label{eq_1_4}
\end{equation}
where \( u_{\mathcal{P}}u^\dagger_{\mathcal{P}}=I \) with \( I \) being the unit matrix. Then, PT symmetry, a combination of the above two symmetries, reads
\begin{equation}
    \begin{gathered}
    (U_{\mathcal{PT}}\mathcal{P}) H^\ast (U_{\mathcal{PT}}\mathcal{P})^{-1}=H
    \end{gathered}\label{eq_1_5}
\end{equation}
in the real space with \( U_{\mathcal{PT}}=U_{\mathcal{T}}U_{\mathcal{P}} \). For the Bloch Hamiltonians, PT symmetry is defined as
\begin{equation}
    \begin{gathered}
    u_{\mathcal{PT}} h^\ast(\bm{k}) (u_{\mathcal{PT}})^{-1}=h(\bm{k})
    \end{gathered}\label{eq_1_6}
\end{equation}
with \( u_{\mathcal{PT}}=u_{\mathcal{P}}u_{\mathcal{T}} \).
\par

A non-Hermitian Hamiltonian with PT symmetry can exhibit EPs, as this symmetry imposes constraints on the complex energy spectra. If \( \phi_{\bm{k}} \) represents a right eigenstate and \( \epsilon_{\bm{k}} \) denotes its corresponding eigenenergy of the Bloch Hamiltonian \( h({\bm{k}}) \), we have the relationship \( h(\bm{k})\phi_{\bm{k}}=\epsilon_{\bm{k}}\phi_{\bm{k}} \). It can be demonstrated that there exists another right eigenstate of \( h(\bm{k}) \) with the energy \( \epsilon^\ast_{\bm{k}} \) and the eigenfunction \( u_{\mathcal{PT}}\phi^\ast_{\bm{k}} \). Thus, in the presence of PT symmetry, if the two states \( \phi_{\bm{k}} \) and \( u_{\mathcal{PT}}\phi^\ast_{\bm{k}} \) are identical, then it follows that \( \epsilon_{\bm{k}}=\epsilon^\ast_{\bm{k}} \), meaning the energy spectrum is real. Conversely, if they are not the same, the eigenenergies appear in pairs of complex conjugates. The critical point separating the real-energy phase from the complex energy phase is referred to as the EP.
\par

On the other hand, the main difficulty in studying how the EP affects the universality of ALTs is to precisely trace the positions of EPs on the complex energy plane. To solve this issue, we further consider imposing the parity-particle-hole (PPH) symmetry. Within the 38-fold symmetry classification, particle-hole symmetry is defined as~\cite{kKohei_prx_2019} 
\begin{equation}
    \begin{gathered}
    U_{\lambda} H^T U^{-1}_{\lambda}=-H
    \end{gathered}\label{eq_1_7}
\end{equation}
in the real space and 
\begin{equation}
    \begin{gathered}
    u_{\lambda} h^T(-\bm{k}) u^{-1}_{\lambda}=-h(\bm{k})
    \end{gathered}\label{eq_1_8}
\end{equation}
in the lattice-momentum space with \( U_{\lambda}U^\dagger_{\lambda}=I \) and \( u_{\lambda}u^\dagger_{\lambda}=I \). As a combination of parity inversion and particle-hole symmetries, PPH symmetry reads
\begin{equation}
    \begin{gathered}
    (U_{\lambda\mathcal{P}}\mathcal{P}) H^T (U_{\lambda\mathcal{P}}\mathcal{P})^{-1}=-H
    \end{gathered}\label{eq_1_9}
\end{equation}
and
\begin{equation}
    \begin{gathered}
    u_{\lambda\mathcal{P}} h^T(\bm{k}) u^{-1}_{\lambda\mathcal{P}}=-h(\bm{k})
    \end{gathered}\label{eq_1_10}
\end{equation}
with \( U_{\lambda\mathcal{P}}=U_{\lambda}U_{\mathcal{P}} \) and \( u_{\lambda\mathcal{P}}=u_{\lambda}u_{\mathcal{P}} \).
\par

Similar to PT symmetry, PPH symmetry also imposes constraints on the complex energy spectrum. The eigenenergies associated with PPH symmetry occur in pairs, represented as \( (-\epsilon, \epsilon) \), which means that the energy spectrum is symmetric around the origin of the complex energy plane. Consequently, if both PT and PPH symmetries are preserved, the corresponding energy spectrum will either be real or purely imaginary, ensuring that the EP is always located at the origin of the complex energy plane. Then, one can easily trace the EP of the non-Hermitian Hamiltonian, even in the presence of randomness.
\par

\subsection{Tight-binding model}
\label{sec2_2}

Inspired by the symmetry analysis above, we consider the following tight-binding model on a cubic lattice of size \( L^3 \): 
\begin{equation}
    \begin{gathered}
    H=\left[ \sum_{\bm{i}}\sum_{\mu=1,2,3}c^\dagger_{\bm{i}+\hat{x}_{\mu}}\left( \dfrac{i\alpha}{2} \Gamma_{\mu} \right) c_{\bm{i}}+h.c. \right] \\
    +\sum_{\bm{i}}c^\dagger_{\bm{i}} (i\kappa \Gamma_4) c_{\bm{i}}+V_{\text{dis}}.
    \end{gathered}\label{eq_1_11}
\end{equation} 
In the above equation, \( c^\dagger_{\bm{i}} \) and \( c_{\bm{i}} \) represent the single-particle creation and annihilation operators at a lattice site \( \bm{i} \). \( \hat{x}_{\mu} \) represents the primitive cell vectors in one direction of the cubic lattice. The parameters \( \alpha \) and \( \kappa \) are positive real numbers. We define the five non-unique Gamma matrices as \( \Gamma_{1,2,3,4,5} = ( \tau_0 \sigma_3, \tau_0 \sigma_1, \tau_1 \sigma_2, \tau_2 \sigma_2, \tau_3 \sigma_2 ) \) and \( \Gamma = \tau_0 \sigma_0 \) with \( \sigma_{0,1,2,3} \) and \( \tau_{0,1,2,3} \) representing the unit matrix and the Pauli matrices, respectively.
\par

The disorder is introduced through \( V_{\text{dis}} \). In the clean limit, the Hamiltonian can be block diagonalized as follows: 
\begin{equation}
\begin{gathered} 
h(\bm{k}) = \alpha (\sin[k_1]\Gamma_1 + \sin[k_2]\Gamma_2 + \sin[k_3]\Gamma_3) + i\kappa\Gamma_4. 
\end{gathered}\label{eq_1_12} 
\end{equation}
This Bloch Hamiltonian \( h(\bm{k}) \)  exhibits PT symmetry because there exists a unitary matrix \( u_{\mathcal{PT}} = \Gamma_{34} \) that satisfies Eq.~\eqref{eq_1_6}. Here, \( \Gamma_{34} = \tau_3 \sigma_0 \), where \( \Gamma_{ij} = [\Gamma_i, \Gamma_j]/(2i) \) is defined generally. Additionally, one can identify \( u_{\lambda\mathcal{P}} = \Gamma_3 \), which fulfills Eq.~\eqref{eq_1_10}, demonstrating that the Bloch Hamiltonian \( h(\bm{k}) \) also preserves PPH symmetry.
\par

\begin{figure*}[htbp]
\includegraphics[width=0.95\textwidth]{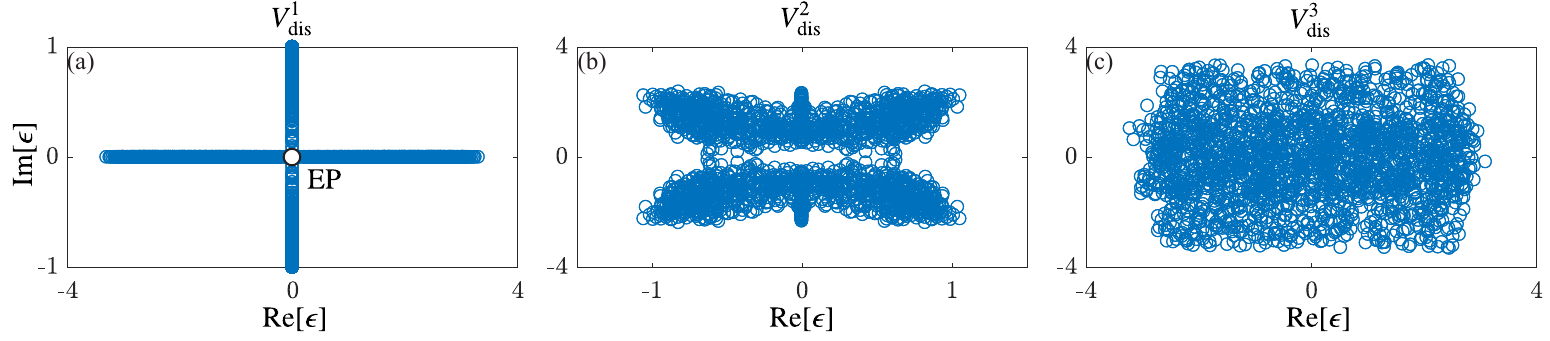}\centering
\caption{(a) The energy spectrum of \( H \) with the disorder term \( V^1_{\text{dis}} \) is shown on the complex energy plane. In this case, we have chosen \( \alpha = 1.0 \) and \( \kappa = 1.0 \). The system size is set to \( L = 6 \), and the random variable \( w_{\bm{i}} \) follows a uniform distribution with a width of \( W = 5.0 \). We used 10 samples to plot the energy spectrum. Due to the symmetries present, the energy spectrum forms a cross shape, with the EP located at \( (0,0) \). (b) and (c) show the same configuration as (a) but for the disorder terms \( V^2_{\text{dis}} \) and \( V^3_{\text{dis}} \), respectively.}
\label{fig1}
\end{figure*}

Let's discuss how to introduce the disorders in this study. We focus on three types of disorders, each leading to different symmetries in the random Hamiltonians. The disorder terms are represented as follows:
\begin{equation}
\begin{gathered} 
V_{\text{dis}}=\left\{
\begin{array}{ccc}
    V^1_{\text{dis}} & = & \sum_{\bm{i}} c^\dagger_{\bm{i}} (w_{\bm{i}}\Gamma_1) c_{\bm{i}} \\
    V^2_{\text{dis}} & = & \sum_{\bm{i}} c^\dagger_{\bm{i}} (i u_{\bm{i}}\Gamma_1) c_{\bm{i}} \\
    V^3_{\text{dis}} & = & \sum_{\bm{i}} c^\dagger_{\bm{i}} [(w_{\bm{i}}+iu_{\bm{i}})\Gamma_0] c_{\bm{i}}
\end{array}
\right..
\end{gathered}\label{eq_1_13} 
\end{equation}
In Eq.~\eqref{eq_1_13}, \( w_{\bm{i}} \) and \( u_{\bm{i}} \) are random numbers that follow specific distributions, which we will discuss later. When the random potential \( V^1_{\text{dis}} \) is present, the disordered Hamiltonian retains both PT and PPH symmetries. Due to these two symmetries, the shape of the energy spectrum appears as a cross on the complex energy plane, with the EP located at the origin. This feature is illustrated in Fig.~\ref{fig1}(a) for one representative example. In contrast, if PT symmetry is broken by the disorders, as seen in the cases of \( V^2_{\text{dis}} \) and \( V^3_{\text{dis}} \), the EP disappears. Notably, while \( V^2_{\text{dis}} \) still preserves PPH symmetry, maintaining the symmetry of the energy spectrum around the origin of the complex energy plane, no symmetry is imposed for \( V^3_{\text{dis}} \). Two representative examples are shown in Fig.~\ref{fig1}(b) and (c).
\par

\subsection{Finite-size scaling analysis}
\label{sec2_3}

We carry on a finite-size scaling analysis of the participation ratio to identify the ALT of \( H \). For a given complex energy \( \epsilon \), the participation ratio is defined as \( p_2 = \left(\sum_{\bm{i}} |\psi_{\bm{i},\epsilon}|^4\right)^{-1} \), where \( \psi_{\bm{i},\epsilon} \) represents the real-space normalized wave function at a site \( \bm{i} \) for the complex energy \( \epsilon \). When an ALT occurs at a critical point \( W = W_c \) with \( W \) quantifying the degree of randomness, the participation ratio is shown to follow a one-parameter scaling law~\cite{cWang_prb_2023}: 
\begin{equation}
\begin{gathered} 
p_2 = L^D f\left(L/\xi\right). 
\end{gathered}\label{eq_1_14} 
\end{equation} 
In this equation, \( D \) denotes the fractal dimension, and \( \xi \) is the correlation length that diverges near the critical disorder \( W_c \) according to a power law, expressed as \( \xi \propto |W - W_c|^{-\nu} \). Here, \( \nu \) is the critical exponent that characterizes the universality class, and \( f(x) \) is the scaling function. 
\par

We numerically calculate the participation ratio using Kwant and Scipy, which are well-established Python packages~\cite{Groth_kwant_2014,scipy}. After obtaining the participation ratio, we can determine the critical disorder \( W_c \), the critical exponent \( \nu \), the fractal dimension \( D \), and the scaling function \( f(x) \) by using a chi-square fitting~\cite{press1996numerical}. We assess the goodness-of-fit, denoted as \( Q \), to evaluate whether the chi-square fit is acceptable, ensuring that \( Q > 10^{-3} \), which exceeds the commonly accepted lower limit for a good fit~\cite{press1996numerical}. This criterion has been met in all the numerical fittings in this paper. 
\par

From the scaling analysis, we can calculate the renormalized participation ratio, defined as \( Y_L = L^{-D} p_2 \). At the critical disorder, the renormalized participation ratio is independent of size. In the delocalization phase, it increases with size \( L \), while in the localization phase, it decreases. This behavior allows us to directly observe the ALT through a plot of the renormalized participation ratio \( Y_L \)~\cite{cWang_prb_2023}. 
\par

Here are some comments regarding our finite-size scaling analysis. Firstly, we cannot diagonalize the Hamiltonian exactly at the EP. Therefore, we examine the nearest-neighbor state of the EP and investigate the ALT, which becomes closer to the EP as the system size \( L \) increases. We argue that its criticality is identical to that of the EP in the thermodynamic limit~\cite{cWang_prb_2023}. Secondly, we have ignored the irrelevant scaling variable in our analysis, as this term is significantly smaller compared to the relevant scaling variable~\cite{cWang_prl_2025}.
\par

\section{Results}
\label{sec3}

\subsection{Disordered systems with PT symmetry}
\label{sec3_1}

\begin{figure}[htbp]
\includegraphics[width=0.48\textwidth]{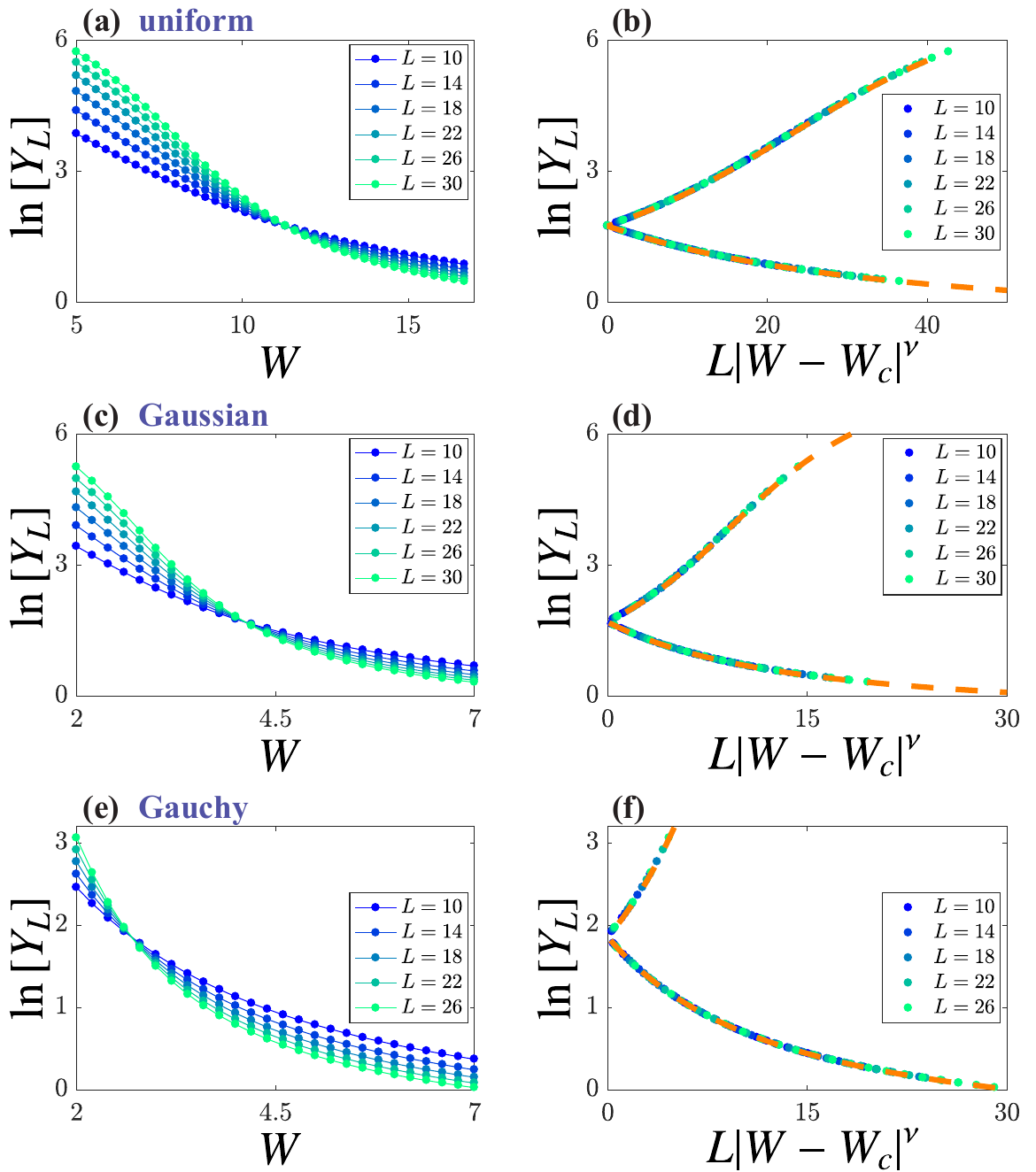}\centering
\caption{
\textbf{ALTs at the EP.}~(a) The natural logarithm of the renormalized participation ratios \( \ln{[Y_L]} \) for \( V^1_{\text{dis}} \) is plotted as a function of disorder \( W \) for various system sizes. We fix the energy at \( \epsilon=0 \), while other parameters are set to \( \alpha=1.0 \) and \( \kappa=1.0 \). Each point is averaged over more than \( 10^2 \) samples. The random variable \( w_{\bm{i}} \) follows a uniform distribution as defined in Eq.~\eqref{eq_1_15}. (b) The scaling function \( \ln{[Y_L]} = \ln{[f(L|W-W_c|^{\nu})]} \) is obtained by merging the data from part (a) with a suitable choice of the critical disorder \( W_c \) and the critical exponent \( \nu \), determined by the finite-size scaling analysis. (c, d) These are the same as (a, b) but for cases where the random variable \( w_{\bm{i}} \) follows a Gaussian distribution. (e, f) Similarly, these represent the same results as (a, b) but for a Cauchy distribution.  
}
\label{fig2}
\end{figure}

\begin{table*}
\caption{
The parameters for the finite-size scaling analysis of ALTs for \( V^1_{\text{dis}} \) are outlined as follows. The first column presents the probability distributions of the random variable \( w_{\bm{i}} \), which can be uniform, Gaussian, or Cauchy. The second column lists the complex energies \( \epsilon \), with \( \epsilon = 0 \) representing the EP. The third column indicates the system length \( L \) used in the finite-size scaling analysis. The fourth column specifies the range of disorder. The fifth, sixth, and seventh columns display the results of the scaling analysis: the critical disorder \( W_c \), the fractal dimension \( D \), and the critical exponent \( \nu \), respectively. The eighth column shows the degrees of freedom (dof) in the chi-square fitting, while the last column indicates the goodness of fit \( Q \).
}
\begin{ruledtabular}
\begin{tabular}{ccccccccc}
\( p(w_{\bm{i}}) \)& \( \epsilon \) & \( L \) & \( W \) & \( W_c \) & \( D \) & \( \nu \) & dof & \( Q \) \\
\hline
uniform & 0    & \( [10,30] \) & \( [5.0,16.7] \) & \( 11.30\pm 0.03 \) & \( 0.607\pm 0.004 \) & \( 1.777\pm 0.006 \) & 233 & 0.1  \\
        & 0.3  & \( [14,30] \) & \( [5.6,16.1] \) & \( 11.43\pm 0.04 \) & \( 0.519\pm 0.007 \) & \( 1.85\pm 0.01 \)   & 173 & 0.1  \\
        & 0.3i & \( [10,26] \) & \( [6.8,15.2] \) & \( 12.1\pm 0.2 \)   & \( 0.48\pm 0.03 \)   & \( 1.86\pm 0.06 \)   & 138 & 0.03 \\
\hline
Gaussian & 0   & \( [10,30] \) & \( [2.0,7.0] \) & \( 4.12\pm 0.02 \)   & \( 0.632\pm 0.004 \) & \( 1.776\pm 0.009 \) & 149 & 0.2   \\
Cauchy   & 0   & \( [10,26] \) & \( [2.0,7.0]\)  & \( 2.687\pm 0.009 \) & \( 0.610\pm 0.005 \) & \( 1.77\pm 0.01 \)   & 123 & 0.02  \\
\end{tabular}
\end{ruledtabular}\label{table_1}
\end{table*}

To begin with, we concentrate on the model with the disordered potential \( V^{1}_{\text{dis}} \), where both PT and PPH symmetries are preserved, and the EP is fixed at the origin, as illustrated in Fig.~\ref{fig1}(a). We numerically calculate the disorder-averaged participation ratios by randomly generating the numbers \( w_{\bm{i}} \) that are uniformly distributed within the range of \( [-W/2, W/2] \). Specifically, the probability density of the random variable \( w_{\bm{i}} \) is given by: 
\begin{equation} 
\begin{gathered} 
p(w_{\bm{i}})=
\left\{ \begin{array}{cc} \frac{1}{W} & |w_{\bm{i}}|\leq W/2 \\ 0   & |w_{\bm{i}}| > W/2 \end{array} \right.. 
\end{gathered}\label{eq_1_15} 
\end{equation} 
Such uniformly distributed random variables are commonly employed in the study of localization problems~\cite{bKramer_rpp_1993}.
\par

Let us focus on the EP where \( \epsilon=0 \). Figure~\ref{fig2}(a) shows the natural logarithm of the calculated disorder-average renormalized participation \( \ln{[Y_L]} \) as a function of the disorder strength \( W \) for various system sizes ranging from \( L=10 \) to \( L=30 \). Different sizes of curves cross at a single point at \( W_c=11.30\pm 0.03 \), indicating that an ALT occurs. The data of \( \ln{[Y_L]} \) increases and decreases with the size \( L \) for \( W<W_c \) and \( W>W_c \), respectively. These are typical features for the extended and localized states when the disorder strength is smaller and larger than the critical disorder \( W_c \)~\cite{cWang_prb_2023}.
\par

To clearly demonstrate the goodness of our scaling analysis, we plot \( \ln{[Y_L]} \) as a function of \( x=L|W-W_c|^\nu \) in Fig.~\ref{fig2}(b), where \( \nu = 1.777 \pm 0.006 \). Data from all different sizes converge into a single curve, with the lower and upper branches representing the localized and extended phases, respectively. Such convergence strongly supports the one-parameter scaling hypothesis. Additional fitting parameters are provided in Table~\ref{table_1}.
\par

A fundamental belief in this field is the concept of universality, which asserts that critical exponents depend solely on universal properties such as symmetries and dimensionality, rather than the specific details of the models~\cite{bKramer_rpp_1993}. For instance, it has been demonstrated that the critical exponent \( \nu \) remains unchanged in the 3D Hermitian orthogonal ensemble, when the distributions of random variables are altered~\cite{Slevin_njp_2014}. Therefore, it is crucial to examine whether criticality near the EPs maintains the concept of universality. To this end, we build upon the ideas presented in Ref.~\onlinecite{Slevin_njp_2014} and recalculate the critical exponents of ALTs of \( H \) with \( V^1_{\text{dis}} \) using various types of random variables.
\par

We begin by analyzing the case where the random variable \( w_{\bm{i}} \) follows a Gaussian distribution, characterized by the following probability density function:
\begin{equation} 
\begin{gathered} 
p(w_{\bm{i}})=\dfrac{1}{\sqrt{2\pi W^2}}\exp\left[ -\dfrac{w^2_{\bm{i}}}{2W^2} \right].
\end{gathered}\label{eq_1_16} 
\end{equation}
Here, \( W \) measures the width of the distribution~\cite{wChen_prb_2019}. The calculated renormalized participation ratios are shown in Fig.~\ref{fig2}(c), from which one can see an ALT happens at \( W_c=4.12\pm 0.02 \). The critical disorder in this case is much smaller than that of the uniform distribution, and the finite-size scaling analysis gives the critical exponent \( \nu=1.776\pm 0.009 \); see Fig.~\ref{fig2}(d) for the scaling function. This critical exponent is consistent with that of the uniform distributions; see Table~\ref{table_1}.
\par

\begin{figure}[htbp]
\includegraphics[width=0.48\textwidth]{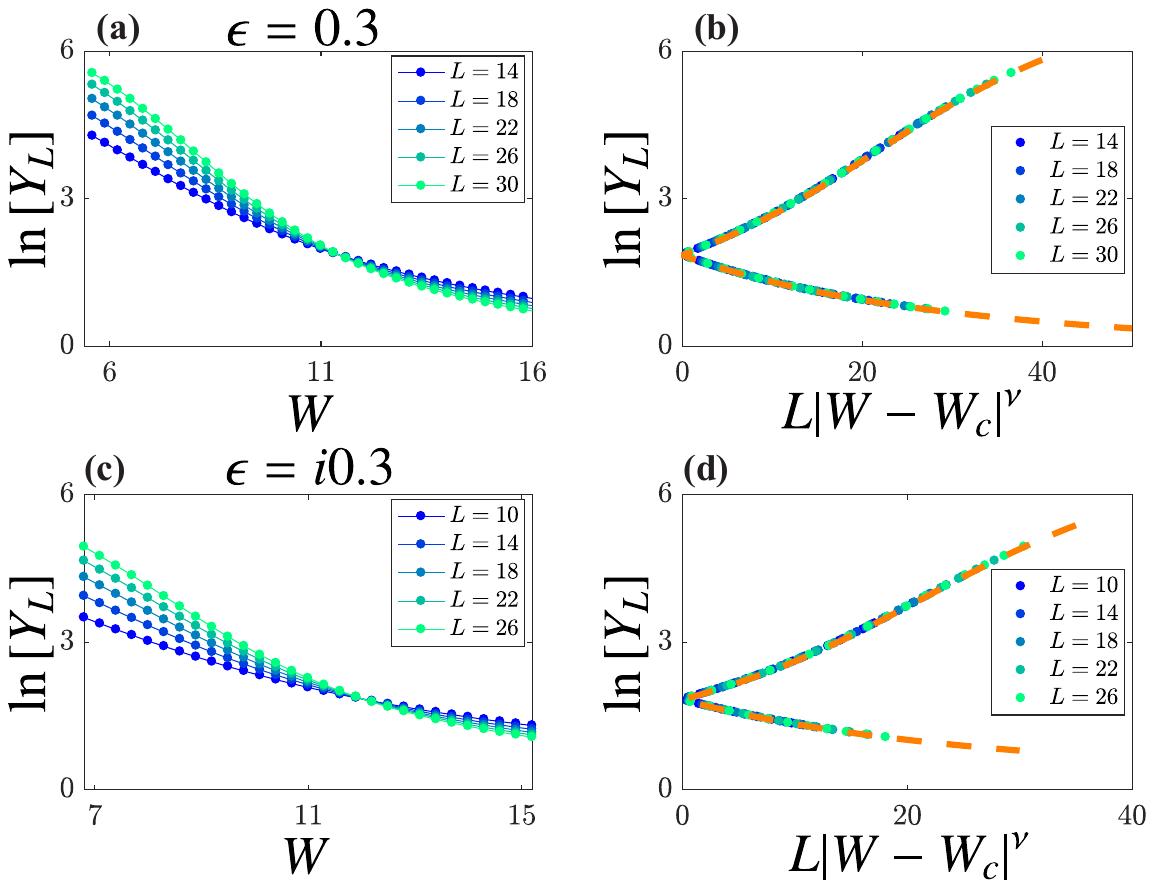}\centering
\caption{
\textbf{ALTs beyond the EP.}~(a) \( \ln{[Y_L]} \) as a function of disorder \( W \) for \( \alpha=1.0 \), \( \kappa=1.0 \), and \( \epsilon=0.3 \) of \( H \) with \( V^1_{\text{dis}} \). The random variable \( w_{\bm{i}} \) follows the uniform distribution defined by Eq.~\eqref{eq_1_15}. (b) The scaling function for \( x=L|W-W_c|^{\nu} \) with \( W_c \) and \( \nu \) being determined by the finite-size scaling analysis. The orange dashed line is the scaling function. (c,d) Same as (a,b) but for \( \epsilon=i0.3 \). 
}
\label{fig3}
\end{figure}

The Cauchy distribution is also utilized to investigate the localization problem. Its probability density function reads
\begin{equation}
    \begin{gathered}
        p(w_{\bm{i}})=\dfrac{W}{\pi (w^2_{\bm{i}}+W^2)}.
    \end{gathered}\label{eq_1_17} 
\end{equation}
Unlike the Gaussian distribution, the Cauchy distribution is classified as a heavy-tailed distribution. This characteristic effectively captures the presence of large fluctuations or rare resonances in disordered media. The obtained renormalized participation ratio \( \ln{[Y_L]} \) as a function of disorder \( W \) is illustrated in Fig.~\ref{fig2}(e). An ALT is observed at \( W_c = 2.687 \pm 0.009 \), where the renormalized participation ratios for various system sizes remain constant. The scaling analysis provides \( \nu = 1.77 \pm 0.01 \), values that align with those derived from uniform and Gaussian distributions, considering numerical errors; see Table~\ref{table_1}. Again, the consistent critical exponents support the concept of universality of ALTs near the EP.
\par

The conventional notion of universality suggests that the critical behaviors of ALTs in a system \( H \) with disorder represented by \( V^1_{\text{dis}} \) are fundamentally the same, as these behaviors depend solely on the system's symmetries and dimensionality~\cite{bKramer_rpp_1993}. While this belief has been confirmed in several disordered systems, recent research indicates that the idea of a universality class extends beyond the concept of a symmetry class~\cite{cWang_prl_2015,ySu_sr_2016,cWang_prb_2017}. One significant implication of these findings is that the introduction of non-trivial topology alters the universality class of ALTs in specific disordered systems, such as in the integer quantum Hall effect~\cite{bHuckestein_rmp_1995}. It is thus crucial to investigate whether the presence of EPs results in a different universality class, since EPs function as topological defects within the parameter space of the system.
\par

To address this issue, we need to examine the critical behavior of ALTs that occur in the same model but go beyond the EP. Therefore, we focus on a representative point \( \epsilon = 0.3 \) along the real axis of the energy spectra of \( H \) with \( V^1_{\text{dis}} \) and calculate the corresponding participation ratio. Using finite-size scaling analysis, we plot the renormalized participation ratio in Fig.~\ref{fig3}(a), which reveals a critical point at \( W_c = 11.43 \pm 0.04 \). Additionally, this analysis yields a critical exponent of \( \nu = 1.85 \pm 0.01 \), which differs from the value observed at the EP. 
\par

Moreover, the fractal dimensions of the critical wave function at these two points (\( \epsilon=0,0.3 \)) differ; see Table~\ref{table_1}. This difference suggests that the critical points have distinct fractal dimensions when approaching or moving away from the EP. Based on these data, the variations in critical exponents are not simply the result of numerical errors. Instead, they indicate that the ALT belongs to different universality classes at the EP compared to when it is away from it. In other words, the emergence of the EP at \( \epsilon=0 \) indeed changes the universality class of the ALT.
\par

To reinforce the conclusion drawn earlier, we present additional evidence by calculating data for a different point at \( \epsilon = i0.3 \) on the complex energy plane, where the renormalized participation ratio and the scaling function are illustrated in Figs.~\ref{fig3}(c) and (d), respectively. Although the strengths of the critical disorder vary slightly, the critical exponents of the delocalization-localization transitions at these two different energy points (\( \epsilon=0.3 \) and \( \epsilon=i0.3 \)) are nearly identical (\( \nu\simeq 1.85 \)). This similarity indicates that the two critical behaviors of ALTs at different energies belong to the same universality class, which is distinct from that observed at the EP.
\par

\begin{figure}[htbp]
\includegraphics[width=0.48\textwidth]{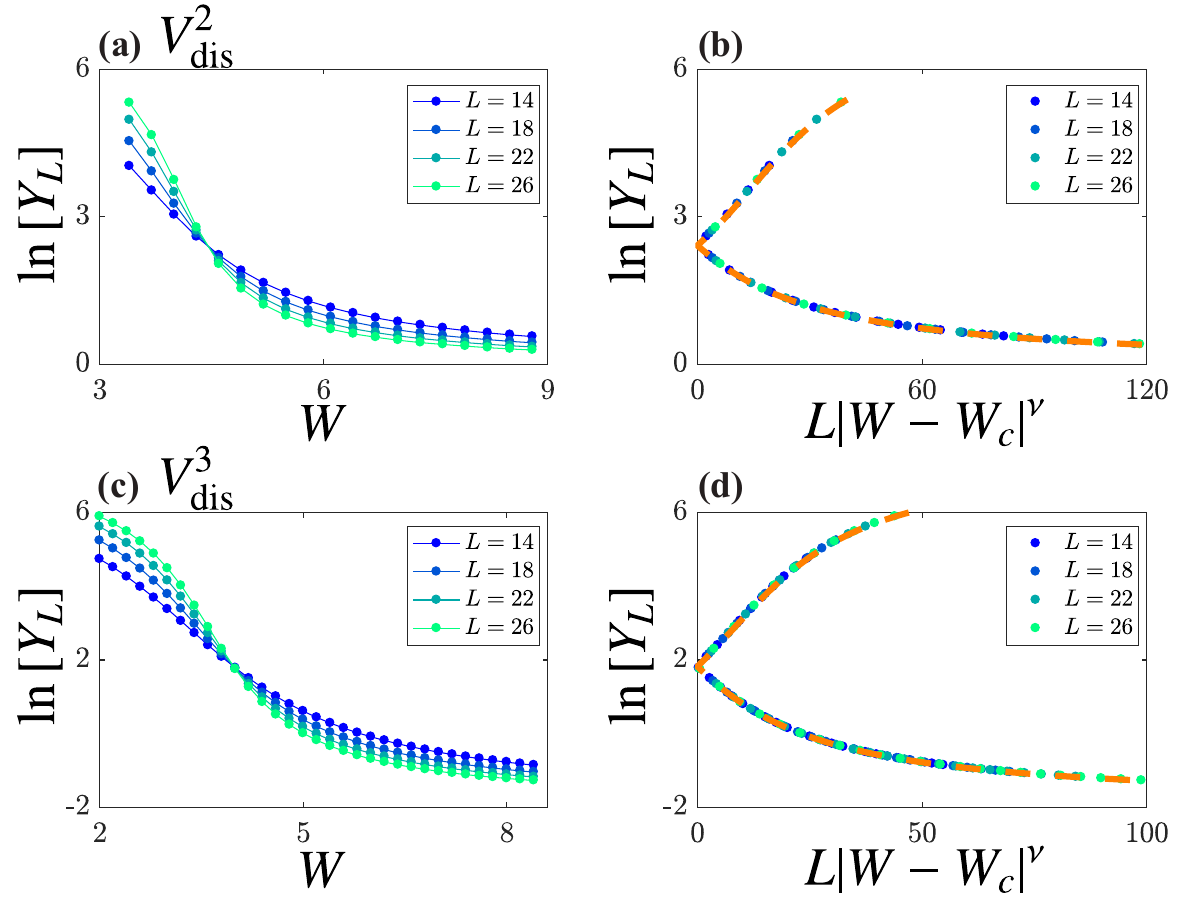}\centering
\caption{
\textbf{ALTs without PT symmetry.}~(a) The plot of \( \ln{[Y_L(W)]} \) is shown for various system sizes \( L \) of \( H \) with \( V^2_{\text{dis}} \), where PT symmetry is broken, and PPH symmetry is preserved. For this analysis, we set the other model parameters to \( \alpha = 1.0 \) and \( \kappa = 1.0 \). The random number \( u_{\bm{i}} \) is uniformly distributed in the range of \( [-W/2,W/2] \). (b) The scaling function is given by \( \ln{[Y_L]} = \ln{[f(x = L|W - W_c|^{\nu})]} \), where the critical disorder \( W_c \) and the critical exponent \( \nu \) are determined through finite-size scaling analysis. The orange dashed line represents the scaling function obtained from a chi-square fitting. (c,d) The results depicted in (a,b) are similar, but this time they correspond to \( H \) with \( V^3_{\text{dis}} \). In this case, both random numbers \( w_{\bm{i}} \) and \( u_{\bm{i}} \) follow a uniform distribution characterized by the probability density function provided in Eq.~\eqref{eq_1_15}.
}
\label{fig4}
\end{figure}

Because of the specific nature of our model \( H \) with \( V^1_{\text{dis}} \), we can transform the real spectrum into an imaginary spectrum by applying a simple transformation: \( H \to iH \). This transformation works because the energy spectrum forms a cross in the complex energy plane, centered at the origin. Unlike EPs, which have topological defect properties, this mathematical transformation should not alter the universality class of the quantum phase transition. As a result, the ALTs at these two different energies (\( \epsilon=0.3,i0.3 \)) have the same critical exponents.
\par

\begin{table*}
\caption{
Fitting parameters for the finite-size scaling analysis of the ALTs in our model \( H \) lacking PT symmetry. The first column gives the form of the disorder potential, where \( V^2_{\text{dis}} \) and \( V^3_{\text{dis}} \) correspond to cases with and without PPH symmetry, respectively. The second column gives the complex energy we considered. The third and fourth columns give the system sizes \( L \) and the range of disorder strength \( W \) used for fitting, respectively. In these cases, the random numbers follow the uniform distribution. The fifth to seventh columns give the critical disorder \( W_c \), fractal dimension \( D \), and critical exponent \( \nu \), respectively. The last two columns give the degrees of freedom (dof) and goodness of fit \( Q \) for the fitting.
}
\begin{ruledtabular}
\begin{tabular}{ccccccccc}
random potential & \( \epsilon \) & \( L \) & \( W \) & \( W_c \) & \( D \) & \( \nu \) & dof & \( Q \) \\
\hline
\( V^2_{\text{dis}} \)   & \( 0.5+0.5i \)    & \( [14,26] \) & \( [3.4,8.8] \) & \( 4.22\pm 0.04 \)   & \( 0.86\pm 0.04 \)   & \( 0.87\pm 0.03 \)     & 69  & 0.01  \\
\( V^3_{\text{dis}} \)   & 0                 & \( [14,26] \) & \( [1.6,8.4] \) & \( 3.975\pm 0.002 \) & \( 0.850\pm 0.003 \) & \( 0.930\pm 0.001 \)   & 133 & 0.01  \\
\end{tabular}
\end{ruledtabular}\label{table_2}
\end{table*}

\subsection{Disordered systems without PT symmetry}
\label{sec3_2}

We further examine the ALT in a disordered non-Hermitian system that lacks PT symmetry but possesses PPH symmetry. This case is associated with the disorder term \( V^2_{\text{dis}} \). The presence of PPH symmetry results in an energy spectrum that is symmetric around the origin of the complex energy plane, meaning the eigenenergies appear in pairs as \( (\epsilon, -\epsilon) \). In contrast, the absence of PT symmetry results in the absence of EP. These characteristics of the energy spectrum are illustrated by one typical example shown in Fig.~\ref{fig1}(b).
\par

Then, we focus on a specific point \( \epsilon= 0.5+i0.5 \) and calculate the participation ratio \( p_2 \) for different system size \( L \) and disorder \( W \) of \( \alpha=1.0 \) and \( \kappa=1.0 \). The random variable \( u_{\bm{i}} \) given in Eq.~\eqref{eq_1_13} follows the uniform distribution Eq.~\eqref{eq_1_15}. The obtained renormalized participation ratios are plotted in Fig.~\ref{fig4}(a), and the corresponding scaling function is given in Fig.~\ref{fig4}(b). The finite-size scaling analysis shows that the critical exponent is \( \nu=0.87\pm 0.03 \), which differs from those for \( V^1_{\text{dis}} \), regardless of the existence of the EP. Therefore, the absence of PT symmetry gives rise to a new universality class. Other parameters of the fittings are given in Table~\ref{table_2}.
\par

To obtain a comprehensive picture of the universality class of our model Eq.~\eqref{eq_1_11}, we consider an additional case with lower symmetry. By selecting a specific random potential, we can simultaneously break PT symmetry and PPH symmetry. This potential is given by \( V^3_{\text{dis}} \) in Eq.~\eqref{eq_1_13}, where the on-site potential is a complex function with both random real and imaginary parts. Here, we also set the random variable \( w_{\bm{i}} \) and \( u_{\bm{i}} \) distributing uniformly in the range of \( [-W/2,W/2] \). 
\par

One representative example of the corresponding energy spectrum is displayed in Fig.~\ref{fig1}(c), from which it can be seen that there are no constraints on the complex energy plane. We then investigate the ALT in this case using numerical methods similar to those employed previously. A representative example of this analysis is shown in Figs.~\ref{fig4}(c) and (d), which considers the ALT for the \( \epsilon=0 \) state.
\par

When disorder increases, a quantum phase transition occurs from extended states to localized states at a critical disorder strength \( W_c=3.975\pm 0.002 \). Such a phase transition can be seen in Fig.~\ref{fig4}(c), which shows the ensemble-averaged natural logarithm of the renormalized participation ratios of different sizes crossing at the critical disorder. Similar to previous observations, the data converge into a single scaling function governed by one parameter \( x = L|W - W_c|^{\nu} \), where \( \nu =0.930\pm 0.001 \); see Fig.~\ref{fig4}(d). The value of this critical exponent appears to be inconsistent with our earlier calculations (see Table~\ref{table_1}), indicating that we have identified a new universality class. 
\par

As a self-consistency check, we also calculate the critical exponents \( \nu \) and the fractal dimensions \( D \) for \( H \) using both \( V^2_{\text{dis}} \) and \( V^3_{\text{dis}} \) under various parameters (results not shown here), e.g., different values of \( \alpha \). Within numerical error, both \( \nu \) and \( D \) obtained for disordered systems of the same symmetries are identical, supporting the universality requirement of the ALT. 
\par

\section{Discussion}
\label{sec4}

We would like to make a few remarks before the conclusion. It is widely accepted that symmetry and dimensionality play significant roles in determining the universality of the Anderson criticalities~\cite{bKramer_rpp_1993}. The orthodox random matrix theory highlights three symmetries (time-reversal, particle-hole, and chiral symmetries) in assessing the universality of ALTs in Hermitian systems~\cite{aAltland_prb_1997}. The presence of non-Hermiticity ramifies and unifies the three symmetries, leading to a 38-fold symmetry classification~\cite{kKohei_prx_2019}. It is thus worthwhile to see whether the 38-fold symmetry classification can solely determine the universality classes of ALTs studied in this paper. 
\par

To this end, we determine the symmetry class of our model \( H \) under different types of disordered potentials defined by Eq.~\eqref{eq_1_13} based on the 38-fold symmetry classification. This classification requires determining the invariance under three symmetry operations (time-reversal, particle-hole, and chiral) and their Hermitian-conjugate operations; see Appendix~\ref{app_1} for more details. For example, our model \( H \) with \( V^3_{\text{dis}} \) only preserves the Hermitian-conjugate of time-reversal symmetry and thus belongs to the class AII$^\dagger$ within the 38-fold symmetry classification. 
\par

Table~\ref{table_3} summarizes the critical exponents \( \nu \), which define the universality class of ALTs, for various disorder terms considered in this paper. It also includes the critical exponents \( \nu' \) derived from different models within the same symmetry class, based on the 38-fold symmetry classification. In the absence of PT symmetry and EPs, the critical exponent obtained is consistent with those of models that share the same symmetries within the 38-fold classification. This fact indicates that, in these symmetry classes, the symmetries can exclusively determine the universality class. Interestingly, when incorporating PT symmetry—whether near or beyond the EPs—the calculated critical exponents \( \nu \) differ from those within the same symmetry class. Such observation suggests that the 38-fold symmetry does not solely dictate the universality class in some non-Hermitian systems. Moreover, the presence of EPs, which are topological defects, leads to the emergence of new universality classes.
\par

\begin{table}
\caption{
Critical exponents \( \nu \) obtained in this paper and \( \nu' \) from the calculations of the same symmetry class based on the 38-fold symmetry classification. The critical exponents for some non-Hermitian symmetry classes have not been determined. Therefore, we identified these critical exponents by mapping the universality classes of the non-Hermitian ALT to those of the Hermitian counterpart with different symmetries, according to a method described in Ref.~\onlinecite{Luo_prresearch_2022}. These critical exponents are marked with an asterisk.
}
\begin{ruledtabular}
\begin{tabular}{ccccc}
                         & \( \epsilon \) & \( \nu \)             & symmetry class  & \( \nu' \) \\
\hline
\( V^1_{\text{dis}} \)   & 0              & \( 1.777\pm 0.006 \)  & AIII+S\(_-\)    & 1.43*\cite{Luo_prresearch_2022}      \\
                         & 0.3            & \( 1.85\pm 0.01 \)    & AIII+S\(_-\)    & 1.43*\cite{Luo_prresearch_2022}      \\
                         & \( 0.3i \)     & \( 1.86\pm 0.06 \)    & AIII+S\(_-\)    & 1.43*\cite{Luo_prresearch_2022}      \\
\hline
\( V^2_{\text{dis}} \)   & \( 0.5+0.5i \) & \( 0.87\pm 0.03 \)    & AII+S\(_+\)     & 0.8745*\cite{Luo_prresearch_2022}    \\
\hline
\( V^3_{\text{dis}} \)   & 0              & \( 0.930\pm 0.001 \)  & AII\(^\dagger\) & 0.903\cite{yAsada_jps_2005}      \\
\end{tabular}
\end{ruledtabular}\label{table_3}
\end{table}

\section{Conclusion}
\label{sec5}

In conclusion, we have explored the universality classes of ALTs in 3D non-Hermitian systems that exhibit PT symmetry. For systems with both PT and PPH symmetries, the energy spectrum on the complex energy plane displays a cross pattern, with the center being an EP that separates the real and complex energy spectra. As disorder increases, states beyond the EP can undergo an ALT characterized by a universal critical exponent \( \nu = 1.85 \). Notably, we have identified a distinct universality class near the EP, characterized by a critical exponent of \( \nu = 1.77 \). This critical exponent remains consistent across various random distributions, including uniform, Gaussian, and Cauchy distributions. We also examined the ALTs in non-Hermitian models lacking PT symmetry, observing that their critical behaviors differ from those involving EPs. Finally, by comparing our findings with studies of ALTs in other non-Hermitian systems, we conclude that the universality class of disordered non-Hermitian systems with PT symmetry and EPs cannot be fully determined solely by the symmetries outlined in the 38-fold symmetry classification.
\par

\begin{acknowledgments}
This work is supported by the National Natural Science Foundation of China (Grants Nos. 12574023 and 12374122). XRW acknowledges the support from the University Development Fund of the Chinese University of Hong Kong, Shenzhen.
\end{acknowledgments}

\section*{Data Availability Statement}

The raw data for all the figures in this work are openly available in a public repository. The custom codes for calculating and analyzing data are available from the corresponding authors upon reasonable request.

\appendix

\section{The 38-fold Symmetry classification}
\label{app_1}

Here, we briefly introduce the 38-fold symmetry classification based on time-reversal symmetry (TRS), particle-hole symmetry (PHS), chiral symmetry (CS), and their conjugate symmetries. We have defined time-reversal and particle-hole symmetries in the main text. Due to the non-Hermitian nature, one must also consider their conjugate symmetries when classifying the symmetry class of a non-Hermitian Hamiltonian. The conjugate symmetry of time-reversal symmetry (denoted as TRS\( ^\dagger \)) is defined as
\begin{equation}
    \begin{gathered}
        U_{\mathcal{P}'}H^T U^{-1}_{\mathcal{P}'}=H
    \end{gathered}\label{app_eq_1_1}
\end{equation}
in the real space and 
\begin{equation}
    \begin{gathered}
        u_{\mathcal{P}'}h^T(-\bm{k}) u^{-1}_{\mathcal{P}'}=h(\bm{k})
    \end{gathered}\label{app_eq_1_2}
\end{equation}
in the lattice-momentum space. The conjugate symmetry of particle-hole symmetry (denoted as PHS\( ^\dagger \)) is defined as
\begin{equation}
    \begin{gathered}
        U_{\mathcal{T}'}H^\ast U^{-1}_{\mathcal{T}'}=-H
    \end{gathered}\label{app_eq_1_3}
\end{equation}
and
\begin{equation}
    \begin{gathered}
        u_{\mathcal{T}'}h^\ast(-\bm{k}) u^{-1}_{\mathcal{T}'}=-h(\bm{k})
    \end{gathered}\label{app_eq_1_4}
\end{equation}
in the real and lattice-momentum spaces, respectively. Here, we utilize the non-Hermitian characteristics of the Hamiltonian (\( H^\ast \neq H^T \)) to generalize TRS and PHS. In fact, the conjugation symmetry of PHS is consistent with TRS. This conclusion can be seen by performing a transformation \( H \to iH \). 
\par

Besides, CS is defined as 
\begin{equation}
    \begin{gathered}
        U_{\Gamma} H^\dagger U_{\Gamma}=-H
    \end{gathered}\label{app_eq_1_5}
\end{equation}
in the real space and 
\begin{equation}
    \begin{gathered}
        u_{\Gamma} h^\dagger(\bm{k}) u_{\Gamma}=-h(\bm{k})
    \end{gathered}\label{app_eq_1_6}
\end{equation}
in the lattice-momentum space. The conjugate symmetry of CS is referred to as sub-lattice symmetry (SLS), which reads
\begin{equation}
    \begin{gathered}
        U_{\mathbb{S}} H U_{\mathbb{S}}=-H
    \end{gathered}\label{app_eq_1_7}
\end{equation}
in the real space and
\begin{equation}
    \begin{gathered}
        U_{\mathbb{S}} h(\bm{k}) U_{\mathbb{S}}=-h(\bm{k})
    \end{gathered}\label{app_eq_1_8}
\end{equation}
in the lattice-momentum space. Within the 38-fold symmetry classification, the symmetry class of a non-Hermitian Hamiltonian is determined by the presence or absence of the above symmetries, and the details of this symmetry classification can be seen in Ref.~\onlinecite{kKohei_prx_2019}.
\par

Here, we directly give the analysis of our model Eq.~\eqref{eq_1_11} in the presence of the disordered potential Eq.~\eqref{eq_1_13}. For \( V^1_{\text{dis}} \), TRS, PHS, and TRS\( ^\dagger \) are broken, while CS and SLS are preserved. Indeed, one can find \( U_{\Gamma}=\tau_2\sigma_2 I \) with \( I \) being the unit matrix of the same dimension as the Hamiltonian and \( U_{\mathbb{S}}=\tau_3\sigma_2 I \) to satisfy CS and SLS. The symmetry operators are anti-commutative, i.e., \( \{ U_{\Gamma},U_{\mathbb{S}}\}=0 \), indicating it belongs to class AIII+S\(_-\). 
\par

Likewise, for \( V^2_{\text{dis}} \), TRS is preserved with a time-reversal operator being \( U_{\mathcal{T}}=\tau_1\sigma_2 I \) and \( U_{\mathcal{T}}U^{\ast}_{\mathcal{T}}=-1 \). SLS is also preserved with \( U_{\mathbb{S}}=\tau_3\sigma_2 I \). The remaining symmetries are not preserved for \( V^2_{\text{dis}} \). In the presence of the two symmetries, \( H \) with \( V^2_{\text{dis}} \) belongs to class AII+S\(_+\). Finally, for \( V^3_{\text{dis}} \), the only symmetry is TRS\( ^\dagger \), with the symmetry operator being \( U_{\mathcal{P}'}=\tau_0\sigma_2 I \). The symmetry operator of TRS\( ^\dagger \) satisfy \( U_{\mathcal{P}'}U^\ast_{\mathcal{P}'}=-1 \), indicating that \( H \) with \( V^3_{\text{dis}} \) belongs to class AII\( ^\dagger \).
\par

\nocite{*}
\bibliography{aipsamp}

\end{document}